\documentclass{iaus}
\usepackage{graphicx}

\usepackage{amsmath}
\usepackage{amssymb}
\usepackage{euscript}

\newcommand{\beq}{\begin{equation}}
\newcommand{\eeq}{\end{equation}}
\newcommand{\bfr}{\mbox{\boldmath ${\rm r}$}}
\newcommand{\bfv}{\mbox{\boldmath $v$}}
\newcommand{\bfX}{\mbox{\boldmath $X$}}
\newcommand{\bfOmega}{\mbox{\boldmath $\Omega$}}

\newcommand{\cendot}{\mbox{\boldmath $\cdot$}}
\newcommand{\ez}{\mbox{{\boldmath $\hat{e}$}}_{Z}}
\title[Binary systems: Outflows \& Periodicities]
{Binary systems: implications for outflows \& periodicities relevant to masers}
\author[Nishant K. Singh \& Avinash A. Deshpande]{Nishant K. Singh$^{1,2}$ \and 
Avinash A. Deshpande$^{1}$}
\affiliation{$^1$Raman Research Institute, C. V. Raman Avenue, Sadashivanagar, 
Bangalore 560080, India\\
emails: {\tt nishant@rri.res.in,\,desh@rri.res.in} \\[\affilskip]
$^2$Joint Astronomy Programme, Indian Institute of Science, Bangalore 560 012, 
India}
\pubyear{2012}
\volume{287}  %% insert here IAU Symposium No.
\pagerange{101--105}
\jname{Cosmic masers - from OH to H0}
\editors{Roy Booth, Liz Humphries \& Wouter Vlemmings, eds.}

\begin{document}
%\pagerange{\pageref{firstpage}--\pageref{lastpage}} \pubyear{2012}

\maketitle

%\label{firstpage}

\begin{abstract}
Bipolar molecular outflows have been observed and studied extensively
in the past, but some recent observations of periodic variations in
maser intensity pose new challenges. Even quasi-periodic maser flares
have been observed and reported in the literature. Motivated 
by these data, we have 
tried to study situations in binary systems with specific attention to 
the two observed features, i.e., the bipolar flows and the variabilities
in the maser intensity. We have studied the evolution of 
spherically symmetric wind from one of the bodies in the binary system, 
in the plane of the binary.
Our approach includes the analytical study of rotating 
flows with numerical computation of streamlines of fluid particles 
using PLUTO code. We present the results of
our findings assuming simple configurations, and discuss the implications.
\keywords{masers, radio lines: general, (stars:) binaries: general, stars: winds, outflows}
\end{abstract}

\firstsection % if your document starts with a section,
              % remove some space above using this command.

\section{Introduction}
\noindent
Bipolar outflows are ubiquitious in nature and presumably thought to 
be associated with star forming regions in molecular clouds. 
Since the discovery of these bipolar outflows 
(\cite[Snell et al. (1980)]{SLP80}) various 
attempts have been made to understand the physical nature of these 
phenomena. 
Such outflows are generally believed to occur around 
%It is generally {\bf believed} that such outflows occur around 
%It is generally agreed that such outflows occur around 
young stellar objects (YSOs) and are thought to have intimate relationship 
with the process of star formation and early stage of stellar evolution 
(see reviews 
\cite[Snell (1983)]{Sne83},
\cite[Lada (1985)]{Lad85},
\cite[Bachiller (1996)]{Bac96}). 
Unanimous view about the 
process of star formation and its early evolution is yet to emerge which 
in turn, inevitably, makes it difficult to have a clear 
%Unanimous opinion about the 
%process of star formation and its early evolution is yet to be made which 
%in turn, inevitably, makes it difficult to have a firm 
understanding of the nature of these bipolar outflows. 
Observations of periodic/quasi-periodic variations in maser intensity
from such regions add further complications
(\cite[Goedhart, Gaylard \& van der Walt (2004)]{GGW04},
\cite[Goedhart et al. (2005)]{GMGW05},
\cite[Goedhart et al. (2009)]{GLGW09},
\cite[van der Walt, Goedhart \& Gaylard (2009)]{WGG09},
\cite[Szymczak et al. (2011)]{SWBL11},
\cite[Araya et al. (2010)]{Ara_etal10}). 

Periodicities observed in the maser light curves are one of the most 
challenging, poorly understood, features.
In the present work, we have tried to demonstrate that a binary system,
consisting of a star and another gravitating object, could be   
a potential candidate to explain {\it together} both of the following observed features
of the maser sources, the bipolarity and the periodicity in the
intensity variations. 
%Periodicities observed in the bipolar outflows are one of the most 
%challenging, yet unexplained, feature.
%In the present work, we have tried to demonstrate that a binary system,
%consisting of a star and another gravitating object, could possibly
%be a potential candidate which can explain both the following observed features
%of the maser observations, the bipolarity and the periodicity in the 
%maser intensity. 

%\section[]{Description of the Envelope\\* Model}
%\section{The Model}
\section{The Model and Simulations}
\noindent
Consider a \emph{binary system} consisting of two bodies, $S$ and $P$,
which are rotating around their common center of mass, $O$, in a plane.
Let $\overline{X}\,\overline{Y}\,\overline{Z}$ be the \emph{inertial}
(fixed) coordinate frame in which the two bodies lie in the
$\overline{X}\,\overline{Y}-$plane with angular velocity vector in
$\overline{Z}-$direction. Let $X\,Y\,Z$ be the \emph{rotating} (corotating)
coordinate frame which rotates with an angular velocity same as that
of the two bodies in the binary system and therefore both the bodies 
appear to be at rest in this frame. The origins of both the coordinate 
frames coincide and are taken to be at the center of mass, $O$, of the
binary system.
The units of various quantities are chosen such that the properties 
of the system depend only on a single parameter. Let the total mass 
(${\cal M}$) of
the primaries ($S$ and $P$) be the unit of mass; the distance between 
them (${\cal D}$) be the unit of distance; and the unit of time be 
chosen in such a way that the angular speed of the primaries, 
denoted by $\Omega$, be unity.
Let $\xi$ be the mass of $P$,
thus mass of $S$ is $(1 - \xi)$. In the rotating reference frame, with
positive $X$ in the direction of the body $P$, the 
coordinates of $P$ and $S$ will be $(1 - \xi, 0)$ and $(-\xi, 0)$ 
respectively. If $(X, Y, 0)$ be the coordinate of an arbitrary point 
$A$, then from Figure~1,
\beq
{\rm r}_{SA} = \lvert \bfr_{SA} \rvert = \sqrt{(X + \xi)^2 + Y^2}\,\,;\quad 
{\rm r}_{PA} = \lvert \bfr_{PA} \rvert = \sqrt{(X - (1 -\xi))^2 + Y^2}
\label{SPcoords}
\eeq
\noindent
The gravitational potential in the corotating frame at the point $A$ 
may be written as,
\beq
\Phi \;=\; -\frac{(1 - \xi)}{{\rm r}_{SA}} - \frac{\xi}{{\rm r}_{PA}}
\label{pot}
\eeq
\noindent
Let us assume that one of the bodies, say $S$, has a 
spherically symmetric wind very near 
to its upper atmosphere, whereas the other body, $P$, is interacting only
gravitationally. If $\bfv(\bfX, \tau)$ be 
the fluid velocity of the wind in the rotating frame
then we may write the Euler equations in rotating frame for steady flow, with
$p$ and $\rho$ as the fluid pressure and density respectively,
as,
\beq
\left( \bfv \cendot \bnabla \right)\bfv \,=\,-\frac{\bnabla p}{\rho}
-\bnabla \Phi - \hat{\bfOmega} \times \left( \hat{\bfOmega} \times \bfX \right)
- 2\hat{\bfOmega} \times \bfv
\label{Euler1}
\eeq
\noindent
where $(\bfX, \tau) \equiv (X, Y, Z, \tau)$ and $\hat{\bfOmega}$ ($= \ez$,
which is the unit vector along $Z\equiv\overline{Z}$) is the angular velocity
of the corotating frame relative to the inertial frame. 

\begin{figure}
\begin{center}
 \includegraphics[scale=0.60]{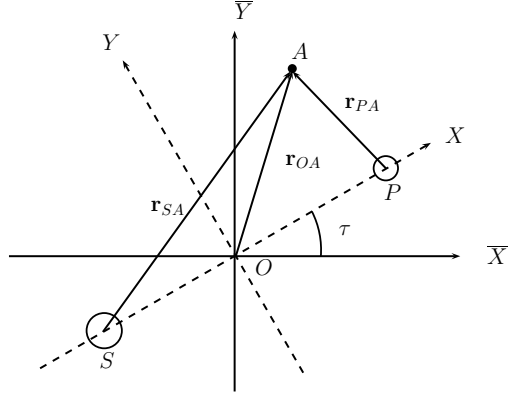}
\caption{The inertial ($\overline{X}\overline{Y}$) and rotating 
($XY$) coordinate frames are shown in the plane of a binary.
%($XY$) coordinate frames are shown for the three--body problem.
The center of mass of the two bodies labelled as $S$ and $P$ is the
origin of both the coordinate frames and is denoted by $O$.}
\end{center}
\end{figure}
\noindent
It can be shown that, 
\beq
(\bfv \cendot \bnabla)\,{\cal B} \;=\; 0
\label{bernth}
\eeq 
\noindent
where
\beq
{\cal B} \;=\; \left( \frac{1}{2}v^2 + \int \frac{d p}{\rho} + 
\Phi_{\rm eff} \right)\;;\qquad \Phi_{\rm eff} \;=\; 
\Phi - \frac{1}{2}\lvert \hat{\bfOmega}\times \bfX \rvert^2
\label{bern_effpot}
\eeq
\noindent
Therefore the quantity, ${\cal B}$, is constant along a particular streamline
for steady flows, although it could be a different constant for different 
streamlines. Noting the fact that the particle paths and streamlines are the 
same for steady flows, we can see that ${\cal B}$ remains the same for a 
particular fluid element as it moves along a particular streamline. Thus, 
for a particular streamline, we may write
\beq
{\cal B} = \frac{1}{2}v^2 + \left(\frac{\gamma}{\gamma - 1}\right) R\, T + 
\Phi_{\rm eff} = \textrm{constant} = C
\label{bernth3}
\eeq
\noindent
where we have used adiabatic equation of state and note that the term
$\int dp / \rho$ appearing in equation \ref{bern_effpot} may be replaced by
specific enthalpy ($w$) for isentropic evolution of fluid element.
$\gamma$ is ratio of specific heats at constant pressure and constant 
volume, $R$ is the gas constant and $T$ is the temperature. As the terms
$v^2 / 2$ and $\{\gamma / (\gamma -1)\}\, R T$ in equation \ref{bernth3}
cannot be negative, we infer from equation \ref{bernth3} that the motion
of a fluid element, and hence the corresponding streamline, is restricted
to the region where $\Phi_{\rm eff} < C$.
%to the region where 
%\beq
%\Phi_{\rm eff} < C
%\eeq
%\section{Numerical simulations using PLUTO code}
\subsection{Numerical simulations using PLUTO code}
\noindent
To study how the spherically symmetric wind from the body
$S$ flows in the presence of another gravitating body $P$ in a binary system,
we use PLUTO code. 
The details of the code may be found in \cite{Mig_etal07} 
(and references therein, and code at http://plutocode.ph.unito.it/). 
We use the hydrodynamic module
%we use PLUTO code {\footnote{See http://plutocode.ph.unito.it/.}.
%The details of the code may be found in \cite{Mig_etal07} 
%(and references therein). We use the hydrodynamic module
of the code and solve the equations in three-dimensional cartesian geometry.
We adapt the code in the corotating frame of the binary in which the two bodies,
$S$ and $P$, appear to be at rest, by adding the necessary body-forces,
namely, coriolis and centrifugal, to the equation of motion. 
To understand the evolution of the wind in the plane of the binary,
we start with a high-pressure circularly symmetric region at the 
location of $S$ and the fluid pressure outside this region being very small,
the fluid particles experience force which is pointed 
radially outward from the location of $S$. 
We have performed simulations for various initial conditions 
(by chosing different values for high/low pressure/density regions) and 
for different values of the parameter $\xi$. Also, the flow need not be steady.
Our simulations are done in two ways: (a) we set the initial conditions
and study the flow. In this case the matter eventually flows out of the 
computational domain as there is no supply of matter from the
location of the body $S$, and the code stops; (b) Having set the initial conditions
we may supply the matter at some arbitrary time intervals. In this case, as the 
%we may supply the matter at some arbitrary interval. In this case, as the 
matter is not completely depleted out of the computational domain, the
code runs for longer time. How quickly the matter is depleted out of the domain 
depends on the initial condition. Thus, we may study the average 
properties of the wind flow (i.e. density, pressure etc.) by plotting 
density/pressure maps at different times.
\begin{figure}
\begin{center}
\includegraphics[scale=0.45]{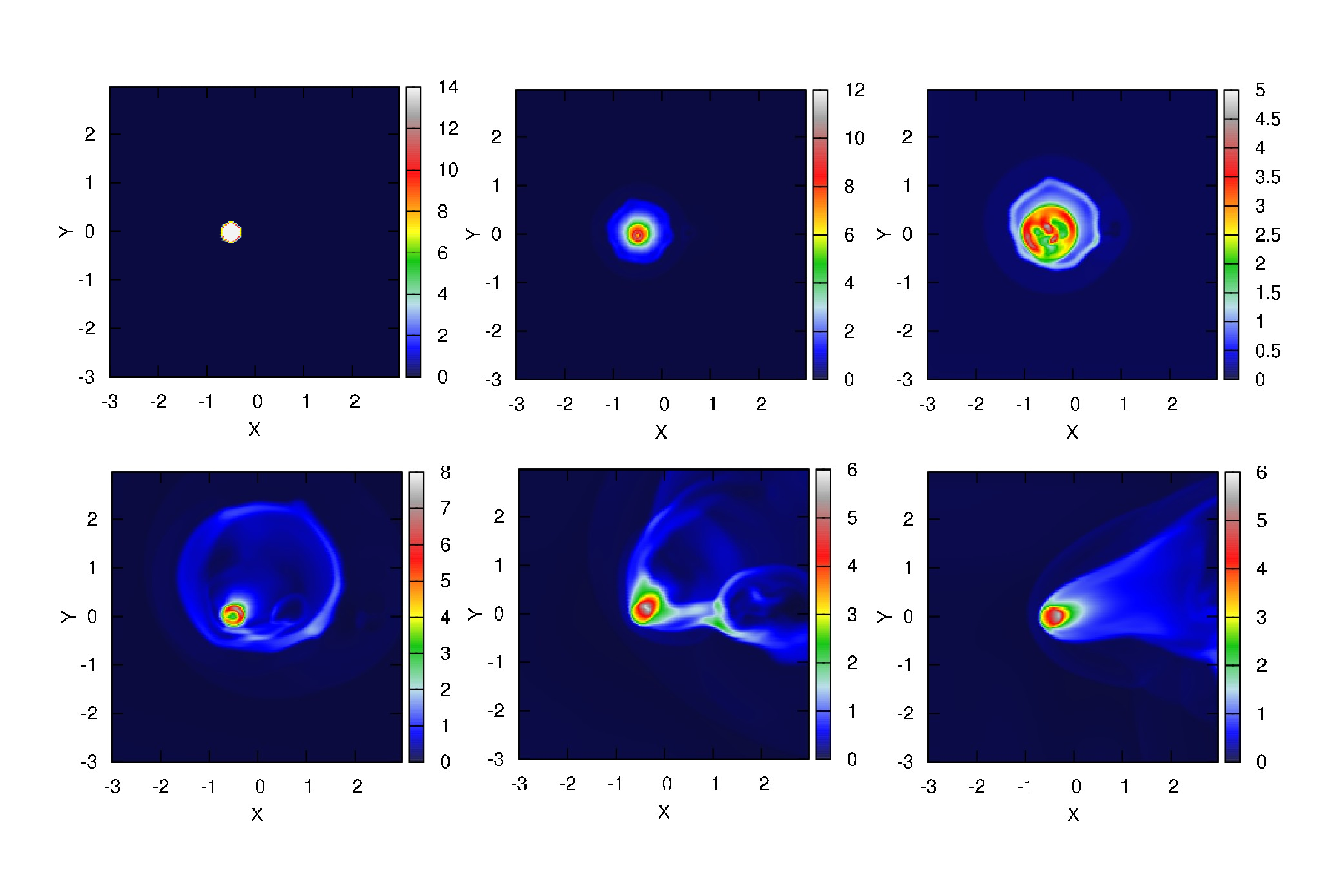}
\caption{Snap-shots of density maps, 
as seen from the corotating frame 
(for $\xi = 0.5$ and with arbitrary supply of matter at $S$), where the bodies
%\caption{Density maps for $\xi = 0.5$ as seen from the corotating frame 
%(with arbitrary supply of matter at $S$), where the bodies
$S$ and $P$ are at $(-0.5, 0)$ and $(0.5, 0)$, respectively. 
Time increases from upper-left (initial time) to lower-right.}
\end{center}
\end{figure}
\subsection{Results from PLUTO simulations and discussion}
We choose to present the result by demonstrating
the evolution of density maps with time as seen in the corotating frame which
is the rest frame of the binary system. This choice of plotting the density
map seems relevant, as ultimately we will be interested in knowing the
distribution of matter in space and its evolution in time to identify the
regions which could potentially be \emph{maser emitting spots}, 
particularly due to relatively high concentration of matter. 
From various panels in Figure~2, we see that the isotropy of the wind
is broken very near to the binary system, as desired for density modulation
seen for a fixed line-of-sight of an inertial observer.
Although it is known that the orbits of a test-body are chaotic in the binary
system, the problem known as \emph{the restricted three body problem} 
(\cite[Poincar\'{e} (1890)]{Poi1890}), it is remarkable to note
that the third test-body
being replaced by the wind (continuous matter) evolves in a similar 
fashion for various initial conditions and for different values of $\xi$
as seen from simulations
(e. g., the average property of the wind, say, density, seems to
evolve in a definitive way). Figure~2 is the result of one of many 
simulations performed to study this problem.
The anisotropy of the wind may be understood to be due to 
the shapes of the isocontours of the effective potential ($\Phi_{\rm eff}$)
which has five Lagrange-points near to the binary, and these isocontours
tend to become Keplerian far away from the binary system. Hence
one may expect that the outflowing matter, which need not escape, 
%one may expect that these outflowing matter, which need not escape, 
settles into Keplerian orbits depending on their initial velocities.
These results prompt us to imagine that on an average, much of the matter
spirals outward with certain pitch-angle, which depends on the intial conditions,
and tends to settle in Keplerian orbits far away from the binary system, 
forming a torus skirting the spiral pattern.  
%These results prompt us to imagine that on an average, much of the matter
%spirals outward with certain pith-angle, which depends on the intial conditions,
%and tends to settle in Keplerian orbits far away from the binary system, 
%and thus forms a torus.  

%\section{Results of variability (light-curve) simulation}
\subsection{Results of variability (light-curve) simulation}
\noindent
We simulate the situation discussed at the end of last subsection and present 
%We simulate the situation discussed at the end of last section and present 
the results by showing light-curves of maser intensities from different locations
as seen along a fixed sight-line of an inertial observer. As it should, the
modulations in the maser intensity will depend on the inclination of 
the binary system with respect to the sky-plane, and so
we show our results for an arbitrarily chosen inclination angle.
Columns of negligible gradients in the sight-line velocity component
with adequate concentration of matter are the most preferred sites of maser
%with enough concentration of matter are the most preferred sites of maser
emission.
\begin{figure}
\begin{center}
\includegraphics[scale=0.5,angle=-90]{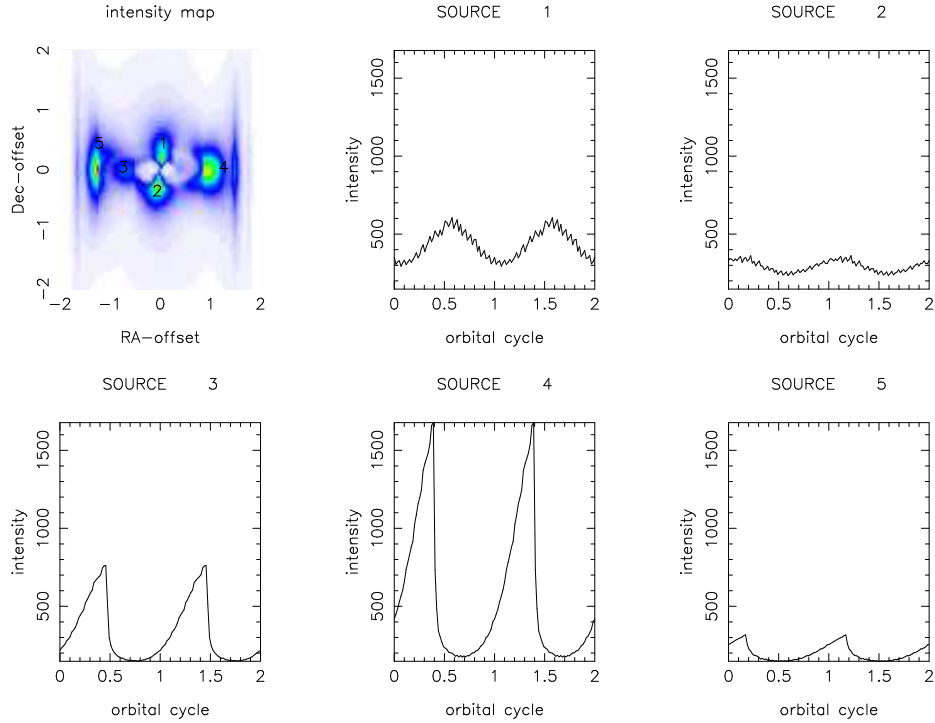} 
\caption{Light-curves (shown as line plots corresponding to the 
locations marked in the color panel, and over two orbital cycles) that an inertial 
observer would see,
as monitored from our simulation of maser emission from flows in binary systems.}
%\caption{Light-curve}
\end{center}
\end{figure}
\section{Summary and Conclusions}
\noindent
We have investigated the flow of the wind from one of the bodies
in a binary system, and tried to understand the plausible
mechanisms for modulations in the maser intensity. Anisotropies seen 
in the rotating frame will have the desired character of bipolar flows and
also will appear to be periodic (with binary-period)
to an inertial observer for relevant sight-lines.
%in the rotating frame will appear to be periodic (with binary-period)
%to an inertial observer for some chosen line-of-sight (except when face-on).
Further, in our model the potential maser spots in the sky-plane do 
\emph{not} move and the \emph{minima} in intensity variation cycle 
repeat at regular intervals of the orbital period. These characteristics
are naturally produced in our model wherein the variabilities 
observed in the maser intensity is due to the density modulation resulting
from the flow in the binary as seen from the simulations.
%Thus, we conclude that the potential maser spots in the sky-plane do 
%\emph{not} move, one of the principal observational facts, 
%and the variabilities observed in the maser intensity is due to the
%density modulation along the corresponding sight-line.  
%density modulation along the chosen sight-line. 
 
{\bf Acknowledgments:}
We thank the IAU, Roy Booth and other symposium organizers for providing
the financial support to one of us (NKS).

%\label{lastpage}

\end{document}